\documentclass[prd,aps,preprint,amsmath,amssymb]{revtex4}
\begin{document}
\tolerance=5000
\def\be{\begin{equation}}
\def\ee{\end{equation}}
\def\bea{\begin{eqnarray}}
\def\eea{\end{eqnarray}}
\def\ii{\'{\i}}
\def\bi{\bigskip}
\def\be{\begin{equation}}
\def\en{\end{equation}}
\def\bq{\begin{eqnarray}}
\def\eq{\end{eqnarray}}
\def\noi{\noindent}
\title{Regular non-twisting S-branes}
\author{Octavio Obreg\'on}
\email{octavio@ifug3.ugto.mx}
\affiliation
   {Instituto de F\'\i sica de la Universidad de Guanajuato \\
                    P.O. Box E-143, 37150 Le\'on Gto., M\'exico}
\author{Hernando Quevedo}
\email{quevedo@physics.ucdavis.edu}
\affiliation{Instituto de Ciencias Nucleares\\
Universidad Nacional Aut\'onoma de M\'exico\\
A.P. 70-543,  M\'exico D.F. 04510, M\'exico\\}
\affiliation{Department of Physics \\
University of California \\
Davis, CA 95616}
\author{Michael P. Ryan}
\email{ryan@nuclecu.unam.mx}
\affiliation{Instituto de Ciencias Nucleares\\
Universidad Nacional Aut\'onoma de M\'exico\\
A.P. 70-543,  M\'exico D.F. 04510, M\'exico\\}
\date{\today}

\begin{abstract}
\setlength{\baselineskip}{.5cm}
We construct a family of time and angular dependent, regular S-brane 
solutions which corresponds to a simple analytical continuation 
of the Zipoy-Voorhees 4-dimensional vacuum spacetime. 
The solutions are asymptotically 
flat and turn out to be free of singularities without requiring a twist in space.
They can be considered as the simplest non-singular generalization
of the singular S0-brane solution. We analyze the properties of a representative
of this family of solutions and show that it resembles to some extent
the asymptotic properties of the regular Kerr S-brane. The R-symmetry corresponds,
however, to the general Lorentzian symmetry.  
Several generalizations of this regular solution
are derived which include a charged S-brane and an additional dilatonic field.

\end{abstract}

\maketitle
\setlength{\baselineskip}{1\baselineskip}

\newpage

\section{INTRODUCTION}
\label{sec:int}

Spacelike branes, or S-branes, are spacelike surfaces similar to ordinary branes 
with the 
special characteristic that one of its transverse dimensions includes time.
They can be interpreted as soliton-like, time dependent field configurations.
In string theory, the study of the potential for the open string tachyon
field \cite{sbranes1} and the search for solutions describing cosmological 
scenarios \cite{sbranes2}
have led to the introduction of S-branes. Their study has received special
attention over the past few years, motivated principally by interest
in understanding the dynamics of time dependent backgrounds in string theory.
In particular, S-branes can be considered as describing the formation and
decay of unstable branes. From the physical point of view, 
the process associated
with the formation and decay of a brane is expected to be smooth. 
In the supergravity approximation, however, it turns out that S-branes 
solutions are plagued by singularities \cite{sbranes1,singsol1}, raising
the question of whether these singularities appear as a result of the
supergravity approximation or due to  other reasons. 
Many other time dependent, asymptotically flat S-branes solutions have 
been analyzed in several works \cite{singsol1,singsol2}, but they  
contain either null singularities or naked timelike singularities
inside internal static regions. Recently, it has been proved that
this singular behavior is an intrinsic property of a large class of
solutions \cite{theosing}. The question arose as to whether, in general,
 there exist 
regular S-brane solutions. This problem is known as the singularity problem
of S-branes.

Among the alternatives suggested to solve the singularity problem
\cite{sbranes1}, the reduction of the R-symmetry of the S-branes 
represents an interesting approach. The idea is that by reducing
the R-symmetry, which represents the symmetry transverse to the 
S-brane worldvolume, the S-brane could be localized in space and
time. In fact, this procedure has been performed recently in 
\cite{wang} and \cite{tasinato}, in the framework of the low-energy
limit of string theory, where it was shown that regular solutions
exist which are less symmetric than the S-brane solutions already
discussed in the literature. The new regular solution 
has been obtained by applying the method of analytical continuation
to the Kerr spacetime and it has been interpreted as a twisting 
S-brane \cite{wang}. The ``rotational" part of the Kerr geometry transforms
into a twist in space and the Kerr angular momentum is reinterpreted
as a twist parameter that determines the global properties of 
the S-brane solution.  
In the limiting case where the Kerr angular parameter 
vanishes, it contains as a special case the singular S0-brane.
Otherwise, the Kerr S-brane solution is regular on the entire manifold
and can also be generalized to include the case of higher dimensions 
\cite{tasinato,luhigh}.
One could then imagine that the twist in space is a necessary
condition in order to get rid of the singularity. We will see 
that this is not necessarily true. In fact, in a recent work \cite{jones}
an analytical continuation of an array of Reissner-Nordstrom black holes
has been interpreted as a regular S-brane configuration. 
In a different approach \cite{lulambda}, regular solutions have been found
in an analytical continuation of an AdS black hole.

In this work, we present a family of regular, non-twisting S-brane solutions
which we derive from the static axisymmetric Zipoy-Voorhees spacetime
by applying the method of analytical continuation. 
In a previous work \cite{samos}, we introduced the horizon method
as a procedure for generating Gowdy cosmological models in General 
Relativity. We have shown \cite{prd}
that the Kerr geometry inside the horizons can be interpreted as
a Gowdy cosmology with topology $S^1\times S^2$. More recently
\cite{zet}, we proposed that the cosmological configurations,
or ordinary D-branes,  
obtained by the horizon method can be used to generate $i$D-branes
which correspond to S-branes. In the present work we generate 
a special type of S-brane solution by applying the method of analytical 
continuation (which would correspond to the $i$-horizon method, 
in our terminology). In fact, we will see that instead of performing
the analytical continuation of the Schwarzschild geometry 
``outside the horizon", one can also perform a similar transformation
``inside the horizon" and this is sufficient to avoid the 
timelike naked singularity present in the non-regular S0-brane solution.

All the solutions we present in this work are
asymptotically flat and can be classified by a real parameter
that determines the explicit time dependence of the 
corresponding metric and curvature of the S-branes. 
In general, these solutions are simpler than 
previous regular S-brane solutions known in the literature 
and can be interpreted as the simplest
regular generalizations of the singular S0-brane solution.
We analyze the main properties of the simplest representative
of this family of regular solutions. We show that the 
regular S0-brane inherits all the symmetry properties of
the original Zipoy-Voorhees solution, but their physical
interpretation is quite different. In particular, we will
see that the $i$-rotation eliminates the original singularity
and that a free parameter entering the metric can even be used
to eliminate the only existent Killing horizon. But if we insist
on preserving the Killing horizon, no changes in the signature
of the metric occur when crossing the horizon, i.e. all the 
coordinates are well behaved on both sides of the horizon. 
In this case, the near horizon limit of the regular S0-brane
solution can be shown to be described by a de Sitter space.
We analyze the R-symmetry of the solution and show that it
corresponds to the general Lorentzian symmetry of a 2-dimensional
conformally flat Euclidean space. In the analysis of the 
asymptotic behavior of the solution we find that the spatial 
asymptote corresponds to a Rindler space with an exponential
expansion in the angular direction, a behavior that coincides
with that of the regular Kerr S-brane. For the temporal 
asymptote we find that 
the spacetime transverse to the worldvolume of the brane  
corresponds to a 2-dimensional Minkowski spacetime with
an exponential expansion in the additional angular direction.
We also derive the charged generalization of the regular 
S0-brane solution and generate a solution which contains
the additional dilatonic field that arises in the low-energy
limit of the IIA string theory. We show that the dilatonic
field modifies the asymptotic temporal behavior of the 
regular S0-brane solution and reduces its Lorentzian R-symmetry
to an $SO(2)$ symmetry.

The paper is organized as follows. After briefly reviewing
the main properties of the 4-dimensional Zipoy-Voorhees spacetime 
in Section \ref{sec:2a}, 
in Section \ref{sec:2b}we use the Schwarzschild metric 
in the Zipoy-Voorhees form ``inside the horizon" to 
derive the simplest regular non-twisting S-brane solution
which we will call the {\it regular S0-brane solution}.
To this end we apply the method of analytical continuation
that has been used in previous works to generate S-brane
solutions. In Section \ref{sec:2c} we study the main 
global properties of this solution. Then, in Section \ref{sec:gen},
we derive different generalized solutions of the regular S0-brane
solution. In particular, we generate and briefly analyze a 
solution that includes, in addition to an electric charge monopole, 
a dilatonic field.  
In Section \ref{sec:3} we present a family of 4-dimensional, 
non-twisting regular S-brane solutions which depend
on two real parameters, $\delta$ and $\mu$, and reduces to the 
special case analyzed in 
Section \ref{sec:2b} when both parameters coincide and 
are taken as $\delta=\mu=1$. 
 Finally, in Section \ref{sec:con}
we comment on our results and 
discuss the possibility of generalizing our results 
to include the case of higher dimensions and additional 
fields of interest in string theory.

\section{A regular non-twisting S-brane}
\label{sec:2}

This Section is devoted to the construction 
 and discussion of the main properties of the simplest, non-twisting, singularity-free
S-brane solution. We first present a brief review of the Zipoy-Voorhees spacetime
which is described by a static, axially symmetric solution of the Einstein vacuum 
field equations and contains the Schwarzschild solution as a special case. 
Then we consider  the special Schwarzschild solution ``inside the horizon" and 
apply the method of analytical continuation to derive the corresponding
S-brane solution, under the condition that it is regular on the entire manifold.

\subsection{The Zipoy-Voorhees spacetime}
\label{sec:2a}

The Zipoy-Voorhees \cite{zv} metric in Lewis-Papapetrou form and 
prolate spheroidal coordinates ($t,x,y,\varphi)$ has the form 
\bea
ds^2 = & - &f dt^2 
+ \sigma^2 f^{-1} (x^2 -1) (1-y^2) d\varphi^2\nonumber \\
       && + \sigma^2 f^{-1} e^{2\gamma} (x^2-y^2)
        \left[ {dx^2\over x^2-1} + {dy^2\over 1-y^2}\right]\ , 
\label{met1}
\eea
with 
\be
f=\left({x-1\over x+1}\right)^\delta \ ,
\qquad e^{2\gamma}= \left({x^2-1\over x^2-y^2}\right)^{\delta^2} \ ,
\ee
where $\sigma$ is a real constant that is used to ``control" the physical
units of the spatial coordinates. 
The constant parameter $\delta$ lies in the range $- \infty
< \delta < + \infty$ with no other restrictions. The degenerate
case $\delta =0$ can be shown to correspond to a flat Minkowski
metric. In general, this spacetime describes a static,
axisymmetric, vacuum gravitational field. Usually it is interpreted
in terms of its multipole moments and corresponds to a non spherically
symmetric mass distribution. The parameter $\delta$ determines all
multipole moments higher than the monopole. In the special
case of $\delta=1$ it reduces to the Schwarzschild metric, as 
can easily be seen by performing the coordinate transformation 
$x=-1 + r/m$ and $y=\cos\theta$, and choosing $\sigma=m$, 
$m$ the Schwarzschild mass.
For this reason, in the general case $(\delta\neq 1)$ one usually
demands that the spatial coordinates lie in the range $x \geq 1$ ($r\geq 2m)$
and $-1\leq y \leq 1$. The 
hypersurface $x=1, \ (r=2m),$ represents a true curvature singularity,
in accordance with the uniqueness theorems of black holes.

If we extend the Zipoy-Voorhees manifold to $x \geq -1$, then 
a second curvature singularity appears at $x=-1$. The sector 
of the manifold contained in the range 
$-1\leq x \leq 1$,
with $-1\leq y \leq 1$, 
can be interpreted as a Gowdy cosmology by means of an appropriate  
redefinition of the coordinates \cite{zet}. This metric is the general 
solution of the $S^1 \times S^2$ Gowdy equations where ? depends only on 
time. This shows that the outer $(x=+1)$ and inner $(x=-1)$ ``horizons"
are actually surfaces of infinite curvature (naked singularities),
but between the ``horizons" this metric 
represents a perfectly viable Gowdy cosmology. 
Notice that in this case, 
an apparent singularity appears at the hypersurface $x=y$ 
which, however, can be removed by means of a suitable coordinate
transformation for any integer values of the parameter $\delta$.
For values within the range $\delta^2 < 3/2$ with $\delta^2\neq 1$, 
the hypersurface $x=y$ corresponds to a true curvature singularity.

There exist certain generalizations of this metric \cite{zvgen} and 
a study of the geodesic motion in the special case $\delta =2$ was performed
in \cite{zvgeo}. Nevertheless, 
the global properties of the Zipoy-Voorhees spacetime have not been analyzed
in detail in the literature except, of course, for the limiting
case  of the Schwarzschild metric ($\delta=1)$. 
This is probably due 
to the fact that only the sector outside the ``horizon" ($x\geq 1$)
 has been  considered of physical interest in the context
of possible applications for describing the exterior gravitational field
of astrophysical objects; however, more general solutions exist in 
the literature (for a review see, for example, \cite{fort}) that are more adequate 
than the Zipoy-Voorhees metric to study non spherically symmetric mass 
distributions and their multipole moments and, consequently are 
more interesting from the point of view of possible astrophysical
applications. Nevertheless, the reinterpretation of
a sector inside the ``horizon" of the Zipoy-Voorhees manifold 
as a cosmological Gowdy 
model \cite{zet} and also as a regular S-brane solution, could
focus new attention on this metric.

\subsection{The regular S0-brane}
\label{sec:2b}

In this subsection we derive a special solution for a non-twisting 
regular S-brane by applying an analytical continuation  
of the Zipoy-Voorhees metric from ``inside the horizon". 
The most general solution that can be obtained in this manner 
will be presented in Section \ref{sec:3}. 
Here we restrict ourselves to the special case where $\delta =1$.
As mentioned before, this limiting case 
corresponds to the Schwarzschild metric with $x= -1 + r/m$,  
$y=\cos\theta$, and $\sigma=m$. By applying an analytical $i$-rotation given
by 
\be
t\to ir\ , \qquad r\to it\ , \qquad \theta \to i\theta\ , \qquad m\to im \ , 
\label{irot0}
\ee 
this black hole solution has been used previously in \cite{s0brane} 
to generate the well-known 4-dimensional S0-brane solution
\be
ds^2 = - \left(1-{2m\over t}\right)^{-1} dt^2  + \left(1-{2m\over t}\right) dr^2 
+ t^2 (d\theta^2 + \sinh^2\theta \ d\varphi^2) \ .\label{ss0brane}
\ee
The disadvantage of this solution is that it possesses a time-like
naked singularity. Therefore it is not appropriate for describing
the formation and decay of unstable branes which is expected to be 
a smooth process free of singularities. 

In the terminology of the horizon method, the $i$-rotation (\ref{irot0})
can be interpreted as an analytical continuation from ``outside the horizon".
What we will show now is that in order to avoid the appearance of singularity
it is sufficient to apply an $i$-rotation from 
``inside the horizon". 

Due to the signature change, between the horizons, $x^2 < 1$, 
the coordinate $x$ becomes timelike and $t$ becomes spacelike. 
Let us define the coordinates 
\be 
x=\cos\tau \ , \qquad t=r 
\ee 
for the cosmological sector of this manifold. Then, from Eq.(\ref{met1}) we obtain
\be
ds^2 = {1-\cos\tau\over 1 + \cos\tau} dr^2 + \sigma^2 (1+\cos\tau)^2 (1-y^2) d\varphi^2
+ \sigma^2 (1+\cos\tau)^2 \left( {dy^2\over 1-y^2} - d\tau^2\right) \ .
\label{zvcosm}
\ee
After an appropriate coordinate transformation \cite{zet}, this metric 
can be interpreted as a 4-dimensional Gowdy cosmological model 
characterized by a big bang singularity at $\cos\tau = -1$. 
We now perform the following $i$-rotation in the cosmological 
sector of the Zipoy-Voorhees metric (\ref{zvcosm}):
\be
\tau \to i \theta \ , \qquad y \to i \frac{\tau}{\sigma} \ , \qquad r\to i r \ .
\label{irot}
\ee
This analytical continuation leads to the following metric
\be
ds^2 = - {\sigma^2 (\cosh\theta + 1)^2 \over \tau^2 + \sigma^2} d\tau^2
+ {\cosh\theta -1 \over \cosh\theta + 1} dr^2 
+ (\cosh\theta + 1)^2[ \sigma^ 2 d\theta ^2 + (\tau^2 + \sigma^2) d\varphi^2 ] \ ,
\label{zv0brane}
\ee
As we discuss in the Conclusions, this process does not guarantee that the 
resultant metric is an axisymmetric time and angular dependent solution of
the vacuum Einstein equations, but one can show that the metric of 
Eq.(\ref{zv0brane}) is indeed a solution. 
An analysis of the corresponding curvature shows that it
is asymptotically flat. Moreover, the Kretschmann scalar  
\be
K = R_{\alpha\beta\gamma\delta}R^{\alpha\beta\gamma\delta}
= {48 \over \sigma^4 (\cosh\theta + 1)^6}
\ee
is perfectly well defined for all values of $\theta$ and does not 
depend on the time coordinate $\tau$. 
One important point about this solution is that it does not
require a ``twist" in space in order to avoid the singularity,
as has been demanded in previous regular solutions 
\cite{wang}. This property makes this metric
the simplest possible regular S-brane solution. From now on,
we will refer to the metric (\ref{zv0brane}) as the {\it 
regular S0-brane solution}. 
 Generalizations
of this solution will be presented in Section \ref{sec:3}.

\section{Global structure} 
\label{sec:2c}

In this Section we investigate the main properties of the regular 
S0-brane derived above. In particular, we are interested in 
the analysis of its isometries, asymptotic behavior and the
underlying R-symmetry.

\subsection{Symmetries}
\label{sec:sym}

Since the spacetime metric (\ref{zv0brane}) does not depend explicitly 
on the spatial coordinates $\varphi$ and $r$ there exist two 
Killing vector fields $K_I = \partial_\varphi$ and $K_{II} = \partial_r$.
$K_I$ describes the axial symmetry and its norm $|K_I| = (\cosh\theta +1)^2
(\sigma^2 + \tau^2)$ is regular on the entire manifold. 
The norm of the second Killing vector 
\be
|K_{II}| =
{\cosh\theta -1\over \cosh\theta +1} 
\ee
vanishes for $\theta = 0$.
This shows  the existence of a Killing horizon at this hypersurface. 
Outside this horizon, the norm of $K_{II}$ is  positive definite, indicating
that the coordinate $r$ is spacelike at all points of the manifold
outside the horizon. 
Notice, however,
that if we preserve the additional parameter $\mu > 1$ (see Section \ref{sec:3}) 
in the case $\delta =1$, the norm of this Killing vector 
$K_{II} = (\mu\cosh\theta -1)/(\mu\cosh\theta +1)$ is positive definite on 
the entire spacetime manifold. In this case no horizon is present. 
Thus, the parameter $\mu$ can be used to eliminate the horizon.

The metric (\ref{zv0brane}) is invariant under the discrete symmetry 
$\theta \to - \theta$, indicating that it is symmetric with respect to 
the hypersurface $\theta =0$. None of the metric functions change their 
sign when passing through the Killing horizon $\theta =0$ so that 
no closed timelike geodesics exist and no Cauchy horizon appears.
An observer traveling along the spatial coordinate $\theta$  
is not affected by the presence of the Killing horizon at $\theta =0$.

\subsection{R-symmetry}
\label{sec:rsym}

The coordinate $\tau$ in (\ref{zv0brane}) is always timelike. The 
explicit dependence on the spacelike coordinate $\theta$ indicates
that the metric can represent a spatially localized source. 
The S-brane worldvolume lies along the spatial direction $r$ so that 
$\varphi$ and $\theta$ are spatial directions transverse to the worldvolume.
While in the singular S0-brane solution (\ref{ss0brane}) and in the regular 
Kerr S-brane solution \cite{wang,tasinato} both transverse directions
$\theta$ and $\varphi$ contain factors which depend explicitly on time. 
In the case of the
regular S0-brane presented here only
the $\varphi$ direction has cosmological expansion. This 
fact represents the main difference in the topology of the 
previous known (singular and regular) S-branes solutions and the 
regular S0-brane solution (\ref{zv0brane}). 

The original singular S0-brane solution 
corresponds to a 4-dimensional manifold of the form $R^{1,1}\times H^2$,
where $H^2$ is a hyperbolic space that determines the R-symmetry. 
Therefore, it possesses a $SO(1,2)$ R-symmetry. In the case of
the regular Kerr S-brane solution, the 4-dimensional manifold 
can be identified as a globally non trivial 
fiber bundle with fiber $H^2$ over the base space $R^{1,1}$. The twisting
in two of the spatial directions, which is the cause of the elimination
of the singularity in this spacetime, induces a non trivial global
topology in the fiber bundle that corresponds to $R^{1,1} \ltimes H^2$,
and the R-symmetry reduces to $SO(2)$ \cite{wang}. 
 
In the regular S0-brane solution (\ref{zv0brane}), the metric of
the transverse directions $\varphi$ and $\theta$ becomes
\be
ds^2_{R-sym} = (\cosh\theta + 1)^2 (\sigma^2 d\theta ^2 +
\tilde\sigma ^ 2 d\varphi^2) \ ,
\ee
where $\tilde\sigma$ is a real constant. This is a conformally flat
2-dimensional Euclidean space which, in general, is invariant with respect to 
transformations belonging to the group $S0(1,3)$. This corresponds to 
the complete Lorentzian symmetry and is therefore
the most general R-symmetry of all known regular solutions.

\subsection{Asymptotic behavior}
\label{sec:asy}

The asymptotic behavior of the regular S0-brane (\ref{zv0brane}) 
is determined by the behavior of the metric as $\theta\to \infty$
and as $\tau \to \infty$. Let us first consider the spatial asymptote.
The only function to be considered is $\cosh\theta$ which behaves
as $\exp(\theta)/2$ for $\theta\to \infty$. The spatial asymptote
is then given by
\be
ds^2_{\theta\to\infty} = {e^{2\theta}\over 4}\left[ 
-{\sigma^2 d\tau^2\over \tau^2+\sigma^2} + \sigma^2 d\theta^2 
+(\tau^2+\sigma^2)d\varphi^2\right] + dr^2 \ .
\label{asym1}
\ee
If we consider, in addition, values of the constant $\sigma$ such that
$\sigma^2<<\tau^2$, this metric reduces to 
\be
ds^2_{\theta\to\infty} = -\xi^2dt^2 + d\xi^2 + {\tau_0^2 \over \sigma^2}
\xi^2 e^{2t} + dr^2 \ ,
\ee
where we have introduced the new timelike coordinate $t=\ln(\tau/\tau_0)$
and the spacelike coordinate $\xi=(\sigma/2)\exp(\theta)$. This metric
corresponds to a Rindler space with an exponential expansion in the
direction of the angle $\varphi$. This coincides exactly with the spatial 
behavior of the 
regular Kerr S-brane solution \cite{wang,tasinato}. 
In the general case of metric (\ref{asym1}) it can 
easily be shown that it corresponds to a Rindler space with an expansion 
in the angular direction given by a hyperbolic cosine function of time.

We now analyze the temporal asymptote. 
In the  asymptotic limit $\tau\to\infty$ we obtain
\be
ds^2_{\tau\to\infty} = 
(\cosh\theta+1)^2(-\sigma^2 dt^2 + \sigma^2 d\theta^2 + \tau_0^2 e^{2t} d\varphi^2 )
+{\cosh\theta -1\over \cosh\theta +1} dr^2  \ ,
\label{asym3} 
\ee
where the time coordinate is given as before by $t=\ln(\tau/\tau_0)$. In the 
special case $r=$ const., this corresponds to a conformal 2-dimensional Minkowski
spacetime with an exponential expansion in the angular direction $\varphi$. 

Although in Section \ref{sec:sym} we have shown that the Killing horizon at 
$\theta =0$ can be removed by means of the free parameter $\mu > 1$, it is
interesting to investigate the limiting case $(\mu=1)$ in which the horizon
is present. From Eq.(\ref{zv0brane}) we can obtain the 3-dimensional metric
in the near horizon limit which can be written as 
\be
ds^2_{nhl} = 4\sigma^2 (-dT^2 + d\theta ^2 + \cosh^2 T \, d\varphi^2) 
+ {1\over 4} (e^\theta -1 )^2 dr^2 \ ,
\label{nhl}
\ee
where we have introduced a new timelike coordinate by means of $\tau = \sigma \sinh T$. 
This shows that near the horizon the regular S0-brane behaves as 
de Sitter space with a time expansion in the angular direction $\varphi$. This
resembles the behavior of the regular Kerr S-brane, but in the present
case crossing the horizon does not imply a change in the signature of 
the metric so that all the coordinates remain well-behaved.

\section{A dilatonic generalization}
\label{sec:gen}

In this Section we will generalize the regular S0-brane solution to include 
the dilaton field which arises in the low-energy limit of IIA string theory.
We first apply a Harrison transformation to the Gowdy cosmological model
presented in Section \ref{sec:2b}. As a result we obtain an electrovacuum
Gowdy solution. The analytic continuation of this 
solution is then interpreted as describing a regular charged S-brane
configuration. Then we make use of a particular symmetry property of the
field equations, which follow from the low-energy action of string theory,
to generate the dilatonic generalization. We briefly discuss the main
properties of this generalized, regular S0-brane solution.

\subsection{A charged regular S0-brane}
\label{sec:charge}

In this Section we derive the charged generalization of the regular S0-brane
presented in \ref{sec:2b}. The first step consists in deriving a solution
of the Einstein-Maxwell field equations which contains the Zipoy-Voorhees 
metric (with $\delta =1)$ in the limiting case of vanishing electromagnetic
field. Clearly, there is in principle an infinite number of possible 
generalizations of a vacuum solution which include an electromagnetic field, 
each one corresponding to different sets of charge multipole moments. 
Here we will present the simplest generalization in which  only 
an additional charge monopole moment is included. In fact, this case 
has been analyzed by Harrison \cite{harrison} who proposed a transformation
which generates electrovacuum solutions from vacuum ones in the following manner. 
Let $f_0$ and $\gamma_0$ represent a vacuum solution with metric (\ref{met1}),
and $f$ and $\gamma$ represent the corresponding electrovacuum solution
with only a charge monopole moment. Then, these two solutions are related by
(see, for instance, \cite{charge} for details of computations):
\be
f= 4 f_0 [ 1 + f_0 + \eta_0 (1-f_0)]^{-2} \ , \quad \gamma=\gamma_0
\ ,\quad \eta_0 = (1 - e^2)^{-1/2} \ , \label{har}
\ee
where $e$ is a real constant which is interpreted as the specific charge,
i.e., the ratio between the net charge and the total mass of the source.
With this transformation one can easily derive the charged generalization
of the Zipoy-Voorhees solution. In the region contained within the ``horizons'',
$-1 \leq x \leq 1$, we proceed as in Section \ref{sec:2b} and introduce 
the coordinates $t=r$ and $x=\cos\tau$ to obtain the metric
\be
ds^2 = {\sin^2\tau\over (\eta_0 + \cos\tau)^2} dr^2 +
\sigma^2 (\eta_0 + \cos\tau)^2\left[ (1-y^2)d\varphi^2 +
{dy^2\over 1-y^2} - d\tau^2\right] \ ,
\label{chargow}
\ee
and the electromagnetic potential 
\be
A = - {e \eta_0\over \eta_0 + \cos\tau} d r \ .
\label{potgow}
\ee
From here we recover the metric (\ref{zvcosm}) in the limiting case $e=0$. 
Consequently, we can interpret this solution as describing a Gowdy cosmology
endowed with a specific electric charge $e$. We now apply the $i$-rotation
(\ref{irot}) on the solution (\ref{chargow}) and introduce the angle coordinate
$\theta$. To obtain a real Maxwell potential we also need to $i$-rotate 
the specific charge, i.e. we change $e\to ie$.  
The resulting metric can be written in the following form:
\be
ds^2 = - {\sigma^2 (\eta+ \cosh\theta )^2 \over \tau^2 + \sigma^2} d\tau^2
+ {\sinh^2\theta \over (\eta+ \cosh\theta)^2} dr^2 
+ (\eta+ \cosh\theta)^2[ \sigma^ 2 d\theta ^2 + (\tau^2 + \sigma^2) d\varphi^2 ] \ ,
\label{zv0charged}
\ee
with the potential 1-form, 
\be
A =  {e \eta\over \eta + \cosh\theta} d r \ , \qquad \eta= (1+e^2)^{-1/2} \ .
\ee
In the limiting case $e=0$, this metric reduces to the regular S0-brane 
presented in Section \ref{sec:2b}. An analysis of the corresponding curvature
shows that the regularity property is not affected by the presence of the 
additional charge. 
As one might expect, the constant 
$e$ determines the charge of the S0-brane.

\subsection{The dilatonic field}
\label{sec:dil}

In this Section we generalize the charged regular S0-brane solution to include
a dilatonic field. Let us consider the Einstein-Maxwell-dilaton action 
\be
S = - \int  d^4 x\, \sqrt{-g} \, [ - R + 2 (\triangle \phi)^2 + e^{-2\alpha\phi} F^2 ] \ ,
\label{dilact}
\ee
where $\phi$ is the dilatonic field and $\alpha$ is the dilatonic coupling 
constant which 
determines the special cases of the theories contained in (\ref{dilact}). Indeed, if 
$\alpha=\sqrt{3}$ one obtains the Kaluza-Klein field equations which result from the
dimensional reduction of the 5-dimensional Einstein vacuum field equations. 
In the special case $\alpha = 1$, the action (\ref{dilact}) coincides with the
low-energy limit of string theory with vanishing dilaton potential. It turns
out that  if we restrict ourselves to spacetimes with two commuting Killing
vector fields, the field equations following from the variation of (\ref{dilact})
possess certain symmetry properties which are very helpful in the search for
new solutions. In particular, one particular symmetry leads to a transformation
that allows us to generate dilatonic solutions
from static electrovacuum solutions. This specific transformation  
can be formulated in the following way. Let 
\be 
f_0\ , \quad \gamma_0\ , \quad  \phi_0=0\ ,\quad F_0=d A_0 
\label{seed}
\ee
represent a particular solution of the field equations following from
the action (\ref{dilact}) with line element (\ref{met1}). Then, a simple
dilatonic generalization of (\ref{seed}) can be obtained by means of the
transformation:
\be
f= (f_0)^{ 1\over 1+\alpha^2}\ , \quad \gamma = {\gamma_0\over 1+\alpha^2} \ ,
\quad A= {A_0\over \sqrt{1+\alpha^2} } \ , \quad
e^{2\phi} = (f_0)^{\alpha\over 1+\alpha^2} \ .
\label{diltrans}
\ee
This transformation can easily be
generalized to the case of time and angular dependent charged solutions with no 
twist such as the solution of Section \ref{sec:charge}. The proof essentially resembles
the procedure we have used in Section \ref{sec:2b} to derive the regular S0-brane
solution. The final result is a new solution of the form:
\bea
ds^2 = &-& (\eta + \cosh\theta )^{2\beta} (\cosh\theta +\tau^2/\sigma^2)^{\alpha^2\beta}
{\sigma^2 d\tau^2\over \tau^2+\sigma^2} + 
\left({\sinh\theta\over \eta + \cosh\theta }\right)^{2\beta}
dr^2 \nonumber \\
&+& (\eta+\cosh\theta)^{2\beta}\left[\sigma^2(\cosh^2\theta + \tau^2/\sigma^2)^{\alpha^2\beta}
d\theta^2 + (\sinh\theta)^{2\alpha^2\beta} (\tau^2+\sigma^2) d\varphi^2\right] \ ,
\label{dilmet}
\eea
\be
e^{2\phi} = \left( {\sinh\theta \over \eta + \cosh\theta}\right)^{2\alpha\beta} \ ,\quad
\beta = {1\over 1+\alpha^2} \ ,
\label{dilaton}
\ee
\be
A= {e\eta\over \sqrt{\beta} (\eta + \cosh\theta)}\, dr \ .  
\label{dilcharge}
\ee
This solution is regular for all values of $\theta$ and $\tau$. There is no change 
in the signature of the metric when crossing the Killing horizon situated at $\theta =0$.
It can be considered as the low-energy limit of a solution of IIA string theory 
(in the Einstein frame). It is therefore possible to construct the corresponding
exact S-brane solution in string theory for which one would expect it to describe
the decay of an unstable D-brane. 

It should be mentioned that the inclusion of the dilatonic field modifies 
some of the global properties of the original regular S0-brane solution.
Unlike the regular S0-brane solution in which only the transverse direction 
$\varphi$ changes in time, the dilatonic field introduces an additional 
time dependence along the spatial direction $\theta$. For the sake of concreteness,
we consider now the special case $\alpha=1$ of the S-brane solution 
(\ref{dilmet})-(\ref{dilcharge}). At spatial infinity 
the asymptotic  behavior of the metric remains unchanged and is described by 
the Rindler space (\ref{asym1}), the electric field vanishes, and the dilatonic
field approaches a constant value. Differences appear in the 
 temporal asymptote which is now given by  
\be
ds^2_{\tau\to\infty} = \sigma^2 (\eta+\cosh\theta )( -4d\tilde\tau ^2 
+ \tilde\tau ^2\,  d\theta^2 + \tilde\tau ^4\, \sinh\theta \, d\varphi^2 )
+ {\sinh\theta\over \eta + \cosh\theta }\,  dr^2 \ ,
\ee
where the new time coordinate $\tilde\tau$ is defined by $\tilde\tau ^2 = \tau/\sigma$. 
Both spatial directions $\theta$ and $\varphi$ present an asymptotic power 
expansion, which is different from the asymptotic exponential expansion
in the $\varphi$-direction of the regular S0-brane solution. 

Another interesting feature of this solution is that the dilatonic field
also modifies the R-symmetry of the original S0-brane solution. Indeed, 
for the special case $\alpha=1$ it is easy to see that the hypersurface
transverse to the worldvolume of the brane is described by the metric
\be
ds^2_{R-sym} = \sigma^2(\eta+\cosh\theta )\left[ \sqrt{\cosh^2\theta+\tau_0^2}\ d\theta^2 + 
\sinh\theta \, d\tilde\varphi ^2\right] \ , 
\label{dilrsym}
\ee
where $\tau_0=\tau/\sigma$ is a constant and $\tilde\varphi = \sqrt{1+\tau_0^2}\ \varphi$.
Unlike the case of the S0-brane solution, this metric is not conformally flat. 
In fact, without the contribution of the conformal factor $(\cosh\theta + \eta)$
the curvature scalar of the metric (\ref{dilrsym}) is given by
\be
R = { (\tau_0^2 -1)\cosh^2\theta - 2\tau_0^2\over 
2\sigma^2\sinh^2\theta\, (\cosh^2\theta + \tau_0^2)^{5/2} } \ ,
\ee
which vanishes (as well as the curvature components) only asymptotically at spatial infinity. 
Since it is possible to introduce a local two-bein for the metric (\ref{dilrsym}) such 
that the local metric is Euclidean, we conclude that underlying symmetry is $SO(2)$. 
This means that the Lorentzian R-symmetry of the original, regular S0-brane solution
becomes reduced to an $SO(2)$ R-symmetry by the presence of the dilatonic field. 

To conclude this Section we would like to mention that it is possible to consider
more general electromagnetic and dilatonic fields. The Harrison transformation 
makes it possible to generate simple charged generalizations from static vacuum
solutions. This simple generalization determines in turn the corresponding
dilatonic field according to the transformation (\ref{diltrans}). In fact, 
all the fields are determined by the seed function $f_0$ of the static vacuum 
solution. 
However, as has been shown in \cite{dilaton}, the field equations which 
follow from the Einstein-Maxwell-dilaton action (\ref{dilact}) possess
more general symmetry properties which can be used to generate electromagnetic
and dilatonic fields in terms of an arbitrary harmonic function. This
harmonic function is completely independent of the value of the seed function
$f_0$. This means that for the regular S0-brane solution presented in 
Section \ref{sec:2b} (or its generalization of the next Section) 
we can generate different dilatonic fields with 
completely different local and global properties. This opens the 
possibility of searching for regular S-brane solutions with 
dilatonic fields with any desired properties.

\section{A family of non-twisting S-branes}
\label{sec:3}

The Zipoy-Voorhees metric (\ref{met1}) is valid for any real values of
the parameter $\delta$. In this Section we generalize the special, regular 
S0-brane solution of Section \ref{sec:2b} to include the case of
an arbitrary value of the parameter $\delta$.

Applying an $i$-rotation ``between the horizons" of the metric (\ref{met1})
and choosing the coordinates in a way similar to that of Section \ref{sec:2b},
 it can be shown that the resulting solution can be written as
\bea
ds^2 = &  \sigma^2&
 f^{-1}e^{2\gamma} (\mu^2\cosh^2\theta +\tau^2/\sigma^2)
\left[ - {d\tau^2\over \tau^2 +\sigma^2 }+ 
{\mu^2\sinh^2\theta \ d\theta^2 \over \mu^2\cosh^2\theta -1}\right] \nonumber \\
&+& f dr^2 + f^{-1} (\mu^2\cosh^2\theta -1) (\tau^2 + \sigma^2) d\varphi^2 \ ,
\label{zvbrane}
\eea
with 
\be 
f=\left( {\mu\cosh\theta -1 \over \mu\cosh\theta +1}\right)^{\delta} \ , 
\quad
e^{2\gamma}= 
\left( {\mu^2\cosh^2\theta - 1\over \mu^2\cosh^2\theta + \tau^2/\sigma^2}\right)^{\delta^2} \ ,
\ee
where $\mu$ is an arbitrary real constant in the range $1<\mu^2<\infty $. 
This is a 4-dimensional axisymmetric, time and angular dependent 
solution of Einstein's vacuum field equations. 
It is regular on the entire manifold
as can be seen from the Kretschmann scalar 
\be
K =  {16 \delta^2\over \sigma^4}  
{ (\mu\cosh\theta - 1)^{-2\delta^2+ 2\delta -2}\over
(\mu\cosh\theta + 1)^{2\delta^2+2\delta + 2} }
(\mu^2\cosh^2\theta + \tau^2/\sigma^2)^{2\delta^2 -3} L(\tau,\theta) \ ,
\ee
where 
\bea 
L(\tau,\theta) = & &3 (\mu\cosh\theta - \delta)^2 (\mu^2 \cosh^2\theta + \tau^2/\sigma^2) 
\nonumber \\
& &+ (\delta^2-1)(1+\tau^2/\sigma^2) 
[ \delta^2 -1 + 3\mu\cosh\theta (\mu\cosh\theta - \delta)]\ .
\eea
For $\mu^2>1$ and any values of the parameter $\delta$ this scalar does not 
diverge at any point of the manifold. The solution is also asymptotically flat. 
For $\delta=1$ we recover the regular S0-brane solution discussed in Section \ref{sec:2b}.
Only in this special case, the parameter $\mu$ can take the degenerate value $\mu =1$.
However, one could also keep this parameter positive and $\mu >1$ and use it to 
eliminate the Killing horizon which appears along the spatial coordinate $\theta$ 
(see Section \ref{sec:2c}).

It is clear that the solution (\ref{zvbrane}) admits a much richer structure than the
regular S0-brane solution (\ref{zv0brane}). For instance, whereas the curvature of 
the regular S0-brane does not depend  explicitly on time, the curvature of the generalized
solution is always time-dependent, and this dependence can be changed arbitrarily  by means 
of the parameter $\delta$. This indicates that the global properties will also 
depend on the parameter $\delta$.

The generalization of this family of regular solutions to include an electromagnetic
field and a dilatonic field is straightforward. The symmetry properties mentioned
in Section \ref{sec:gen} are valid and can be applied in this general case also.

\section{Conclusions}
\label{sec:con}

We have derived a family of 4-dimensional regular S-brane solutions 
that solve the singularity problem of S-branes without requiring 
a twist in space. The simplest representative of this family of 
solutions is obtained by applying the method of analytic 
continuation to the Schwarzschild spacetime, but instead of applying
it from ``outside the horizon", we first ``cross" the horizon 
where the spacetime can be interpreted as a Gowdy cosmology and
then perform the $i$-rotation. This shows that the analytical 
continuation of the Schwarzschild spacetime ``inside the horizon"
can get rid of the singularity without introducing a twist in space.
In the general case, the family
of metrics can be interpreted as a Zipoy-Voorhees regular S-brane 
solution. 

It is important to clarify a point concerning the method used here
and in other works to derive this type of new spacetimes. 
Modern solution generating techniques were developed 
almost thirty years ago and have been extensively used to 
generate solutions of the Einstein-Maxwell field equations
with two Killing vector fields. Simple examples of 
these methods are the Harrison transformation \cite{harrison} as 
given in Eq.(\ref{har}) 
and the dilatonic transformation (\ref{diltrans}). 
All these methods are based
upon the existence of certain continuous deformations 
(Lie transformations) of the field equations which
are the elements of an infinite dimensional group
of transformations, first discovered by 
Geroch \cite{ger1} in the stationary, axisymmetric Einstein 
vacuum field equations. The study of the Geroch group
gave rise to the development of well established methods
that allow us to generate new solutions from  a known seed 
solution.
It is clear that the ``horizon method", the $i$-rotation and other
similar methods cannot be included within the
category of continuous deformations of the field equations
and should be considered at most as ``tricks" that incidentally 
happen to generate new solutions.
Consequently, one can not assure a priori that the application 
of these ``tricks" really lead to new solutions. One always 
should test the resulting metrics with the corresponding
field equations. In fact, in our experience we have found
cases in which these ``tricks" fail to work. Probably, the
reason why they sometimes happen to lead to new solutions
is related to the existence of yet unknown discrete 
symmetries of the field equations.

The S-branes solutions presented here correspond to asymptotically flat,
time and angular dependent backgrounds. They depend on a real parameter
$\delta$, which could be used to ``control" the time and angular dependence
of the  corresponding spacetime metric and curvature, and on an 
additional parameter $\mu$, which in the limiting case $\delta =1$ 
can be used to remove the Killing horizon that appears along one
of the spatial coordinates.

In this work we have analyzed in detail the global properties of 
only the simplest regular S0-brane, i.e., when $\delta=\mu=1$.
This case is relatively simple because the curvature does not
depend explicitly on time. Although this regular solution 
does not have a twist in space, its asymptotic behavior resembles
to some extent that of the twisting Kerr S-brane. 
We have seen that  
the R-symmetry of the regular S0-brane solution
corresponds to that of a 2-dimensional 
conformally flat
space, in contrast to the hyperbolic space R-symmetry of the singular
S0-brane. 
This kind of symmetry reduction has been proposed earlier
as a possible approach to avoid the singularity, and has
been used recently to derive regular twisting S-brane solutions.
Our results show that non-twisting, simpler, regular solutions 
can be obtained when the R-symmetry corresponds to the general
Lorentzian symmetry. However, when we consider the additional
dilatonic field the R-symmetry becomes $SO(2)$ as in the 
case of the twisting Kerr S-brane. This is again in the spirit
of the idea that the reduction of the R-symmetry allows us to
overcome the singularity problem of S-branes. 
Our analysis of the properties of the regular S0-brane solution
and its dilatonic generalization 
shows that it can be used to describe the formation and 
decay of an unstable D-brane.  

The family of regular S-brane solutions derived in Section 
\ref{sec:3} offers several possibilities to continue the
investigation of regular S-brane configurations. 
For instance, 
it would be interesting to study 
the case $\delta = 2$ and $\mu^2 > 1$ for which one can show
that the curvature depends explicitly on time, unlike the
regular S0-brane solution whose curvature depends only
on one spatial coordinate. Also, the corresponding metric
contains an explicit time and angular dependence in both
spatial directions $\theta$ and $\varphi$ which are 
the directions transverse to the worldvolume of the brane.
The asymptotic behavior of 
the corresponding S-brane solution and its dilatonic 
generalization will present 
different possible scenarios that could be of interest, 
especially in the context of the formation and decay of
more general unstable branes. This task is currently under investigation.

\section*{ACKNOWLEDGMENTS}

This work was supported in part  by DGAPA-UNAM grant IN112401,  
CONACyT-Mexico grants 36581-E and 37851-E, grants PROMEP and UG, and by 
U.S. DOE grant DE-FG03-91ER 40674.
H.Q. thanks UC MEXUS-CONACyT (Sabbatical Fellowship Program) 
for support.

\end{document}